\documentclass[%
 reprint,
 amsmath,amssymb,
 aps,
]{revtex4-2}

\usepackage{graphicx}
\usepackage{dcolumn}
\usepackage{bm}

\usepackage{blkarray}
\usepackage{multirow}

\usepackage{soul}
\usepackage{cancel,siunitx}

\begin{document}

\preprint{APS/123-QED}

\title{Measuring nodes' centrality when local and global measures overlap} 
\thanks{Publication rights licensed to APS. This is the authors’ version of the work. It is posted here for your personal use. Not for redistribution. The definitive Version of Record was published in Physical Review E, DOI: \url{https://doi.org/10.1103/PhysRevE.105.044317}.}

\author{Lorenzo Costantini}
 \email{Corresponding author: lorenzo.costantini@polito.it}
\author{Carla Sciarra}%
\author{Luca Ridolfi}
\author{Francesco Laio}
\affiliation{%
 DIATI, Politecnico di Torino, Corso Duca degli Abruzzi 24, 10129, Torino, Italy (IT)
}%

\date{\today}

\begin{abstract}
Centrality metrics aim to identify the most relevant nodes in a network. In literature, a broad set of metrics exists, either measuring local or global centrality characteristics. Nevertheless, when networks exhibit a high spectral gap, the usual global centrality measures typically do not add significant information with respect to the degree, i.e., the simplest local metric. To extract new information from this class of networks, we propose the use of the GENeralized Economic comPlexitY index (GENEPY). Despite its original definition within the economic field, the GENEPY can be easily applied and interpreted on a wide range of networks, characterized by high spectral gap, including monopartite and bipartite networks' systems. Tests on synthetic and real-world networks show that the GENEPY can shed new light about the nodes' centrality, carrying information generally poorly correlated with the nodes' number of direct connections (nodes' degree). 

\end{abstract}

\maketitle

\section{\label{sec:Introduction}Introduction}
In the last few decades, the use of network theory to unravel systems' features has emerged and spread over different disciplines~\cite{newman2018networks,networkBookNewman,BOCCALETTIbook}, with relevant applications in biology~\cite{BioNetwork}, economics~\cite{EconomicNetwork}, epidemics~\cite{EpidemicNetwork}, social sciences~\cite{SocialNetworkBorgatti}, electrical~\cite{pagani2013power,motter2013spontaneous} and computer science engineering~\cite{liu2016modeling,karyotis2016malware}.
A network describes the interactions among elements in a system: the elements are called \emph{nodes}, while the interactions \emph{edges} or \emph{links}~\cite{newman2018networks,networkBookNewman,BOCCALETTIbook}, and for each system a proper matrix representation exists. 

It is crucial to determine the importance of the nodes in a network for understanding the features of the system it represents~\cite{borgatti2005centrality}. Centrality metrics aim to accomplish this task by focusing on the matrix representation of the system. Several centrality measures exist, depending on the chosen criteria to score nodes' importance. Among the most common centrality metrics, one finds the degree and eigenvector centrality~\cite{newman2018networks,networkBookNewman,borgatti2005centrality}. The degree centrality measures the importance of a node through the number of its connections~\cite{newman2018networks,BOCCALETTIbook}, whereas the eigenvector centrality ranks each node accounting for the centrality of the nodes to which it is linked~\cite{newman2018networks,bonacich2007eigenvector}, and thus considering the entire network structure. In light of such difference of perspective about the network topology, the eigenvector centrality is known to be a global measure of centrality, whilst the degree a local one, since its centrality analysis is limited to the nodes' neighborhoods~\cite{newman2018networks,bonacich2007eigenvector,borgatti2005centrality,benzi2015limiting,salavaty2020integrated}. Other global centrality metrics include closeness~\cite{freeman1979centrality}, betweenness~\cite{freeman1977set}, subgraph centrality~\cite{estrada2005subgraph} and total communicability~\cite{benzi2013total}, to cite a few.

Notwithstanding the difference in their rationales, high correlation coefficients among the centrality metrics have been observed in high spectral gap networks~\cite{oldham2019consistency,benzi2013ranking,benzi2015limiting,estrada2006spectral,estrada2006network}, namely those networks presenting a significant difference between the two largest eigenvalues of the matrix representing the system. High spectral gap networks are homogeneous (i.e., with repeated edge patterns), and well-connected (i.e., all networks areas can be easily reached moving from one node to the others)~\cite{tanner1984explicit,estrada2006spectral,alon1986eigenvalues,hoory2006expander,estrada2007topological,milanese2010approximating,jun2010natural}. Several studies have shown that the larger spectral gap, the higher is the Spearman's correlation between both local (hence the degree) and global centrality measures~\cite{oldham2019consistency,benzi2015limiting,li2015correlation}. In such cases, when comparing the resulting rankings, the degree centrality (i.e., the simplest metrics) is a good proxy for the information provided by the global metrics~\cite{benzi2013ranking,li2015correlation}. Therefore, for the networks featuring high spectral gaps, it emerges the need for new global metrics providing additional information about the nodes' centrality beyond the local one of the degree.

Against this background, we propose to extract such information by applying a centrality measure introduced in the field of Economic Complexity, namely the Generalized Economic Complexity index (GENEPY)~\cite{sciarra2020reconciling}. 
The GENEPY metric proved capable of broadening the narrow information that the degree centrality provides about the analysis of the export capacity of countries in the international trade of commodities~\cite{hidalgo2009building,tacchella2012new,sciarra2020reconciling}. This feature suggests that the GENEPY framework can be extended to general network systems characterized by high spectral gaps, in order to gain complementary information with respect to the degree centrality. To this aim, after having provided further proof regarding the correlation among common centrality metrics, we present the GENEPY and its application to high spectral gap networks. We refer to bipartite networks in our main results, but these are easily extended also to monopartite ones. In testing the performances of the GENEPY on a wide set of networks, ranging from artificially generated to real-world ones, we show that the information provided by the GENEPY is less related to the degree of the nodes than other metrics, thus shedding new lights about the nodes' centrality and overcoming the limitations of other metrics.

\section{Mathematical framework}
\subsection{Network science definitions and centrality measures}

Networks (or graphs) can be mainly classified into two categories: monopartite and bipartite networks.
In \emph{monopartite} networks, there exist only one type of interacting entities, and the system can either be undirected or directed depending on the presence of directions in the edges. For example, the network representation of the marriage relationships among the Renaissance families in Florence is undirected, since marriage is mutual~\cite{newman2018networks}; whereas, a directed network suitably represents the citation space of scientific literature~\cite{newman2018networks}. Monopartite networks are described by a square matrix \textbf{A}, called \emph{adjacency} matrix, which is symmetric if the network is undirected, and asymmetric when the graph is directed~\cite{newman2018networks}. The dimension of \textbf{A} depends on the number of nodes in the network, and its entries $A_{ij}$ describe the existence of a link between the nodes $i$ and $j$, in either logical information ($1$ or $0$) or in weighted form. 
Instead, \emph{bipartite} networks describe systems with two different kinds of interacting entities, and  the nodes can be divided into two disjoint subsets $U$ and $P$. In such a system, the interactions only occur among the elements of different sets~\cite{newman2018networks,TwoModeData}. The \emph{incidence} matrix \textbf{B} describing such systems is rectangular with dimension $N_U\times N_P$, i.e., the number of elements is $U$ and $P$, respectively. The element $B_{ij}$ is $1$ when there exists a link between the node $i$ in $U$ and $j$ in $P$, and $0$ otherwise~\cite{newman2018networks,TwoModeData}. Examples of bipartite systems are the international trade of goods, where the interacting entities are the exporting countries and corresponding export baskets~\cite{hidalgo2009building,tacchella2012new}, and the plant-pollinator ecosystems, identifying the two sets in the interacting species of flora and fauna~\cite{hegland2010monitor}. 

Let $G(V,E)$ be a graph where $V$ and $E$ are the sets of the nodes and edges, respectively (if $G$ is bipartite, then $V=U\cup P$). The neighbors of a node $i$ are all the nodes to which $i$ is connected~\cite{newman2018networks}. The graph is defined as \emph{connected} if, independently from the starting node, it is possible to reach all the other nodes in the network~\cite{benzi2013ranking,benzi2015limiting}. A \emph{walk} of length $m$ is a set of nodes $v_1$, $v_2$, \dots, $v_{m+1}$ such that for any integer value of $r$ between $1$ and $m$ the nodes $v_r$ and $v_{r+1}$ are connected by a (directed) link. A \emph{closed walk} is a walk where $v_1=v_{m+1}$. A walk without repeated nodes is a \emph{path}, and a closed walk without repeated nodes (except the starting and arriving node) is a \emph{cycle}~\cite{benzi2013ranking,benzi2013total,benzi2015limiting}. $G'(V',E')$ is a \emph{subgraph} of the graph $G$ if $V'\subseteq V$ and $E'\subseteq E$~\cite{estrada2005subgraph}.

To discuss about the correlations among the centrality metrics (including the degree, eigenvector, closeness, betweenness, subgraph centrality and total communicability) here follows a brief description and interpretation of those centrality metrics. For the sake of conciseness, the description is limited to the general metrics' rationales and we refer the reader to~\cite{newman2018networks,networkBookNewman} for further mathematical details (also concerning to their tailoring to the different network types). 
The degree centrality (D) of a node measures its number of connections~\cite{newman2018networks}, computed as the sum of all the entries of the matrix referring to that node. 
The eigenvector centrality (E) scores the nodes according to the number and importance of the nodes to which they are connected. In mathematical terms, the eigenvector centrality equals computing the first eigenvector (the one associated with the largest eigenvalue) of the matrix describing the network at hand~\cite{bonacich2007eigenvector,newman2018networks,TwoModeData,estrada2005subgraph}. 
The closeness centrality (C) measures the mean topological distance from one node to the others~\cite{newman2018networks,freeman1979centrality,li2015correlation,borgatti2005centrality}.
The betweenness centrality (B) measures the times a node appears along the shortest paths among vertices~\cite{newman2018networks,freeman1977set,salavaty2020integrated}. Supposing a flow of information moving through the system, the betweenness centrality defines the amount of information that moves through a node~\cite{li2015correlation,borgatti2005centrality}. 
The subgraph centrality (SG) of the nodes in a network is defined as the weighted sum of all the closed walks (of any length) starting and ending in the same node~\cite{estrada2005subgraph} and, the shorter the closed walk, the higher is the associated weight. The subgraph centrality scores the nodes considering their participation in all the subgraphs of the considered network~\cite{estrada2005subgraph}. 
The total communicability centrality (TC) of a node $i$ is defined as the row-sum of the $i$-th row of the exponential of the matrix describing the considered network~\cite{benzi2013total}. Under the assumption of information spreading, this metric quantifies how well a node spreads information to any other node in the network. 
All the aforementioned metrics, exception made for the degree, consider the entire network structure in the assignment of a centrality score, and hence they are considered as global centrality measures~\cite{BONACICH1991155,bonacich2007eigenvector,benzi2015limiting,farahat2006authority,li2015correlation,borgatti2005centrality,benzi2013total,salavaty2020integrated}.

The spectral gap, which we will show determines the correlations among these centrality metrics, can be computed for all network kinds. Nevertheless, while the computation of the eigenvalues is straightforward for the square matrices describing undirected networks~\cite{benzi2013ranking}, rectangular matrices need some details. In particular, considering a bipartite network and its incidence matrix \textbf{B}, to compute the spectral gap entails considering the existing relationship between the eigenvalues $\lambda$ of the matrices $\textbf{BB}^\text{T}$ and $\textbf{B}^\text{T}\textbf{B}$. The matrices have dimensions $N_U\times N_U$ and $N_P\times N_P$, respectively, and they represent the monopartite projections of the bipartite system. The spectra (i.e., set of all eigenvalues) of these two matrices coincide up to the smallest value $n$ between $N_U$ and $N_P$ ~\cite{newman2018networks} (the remaining ones are zero). By ordering the eigenvalues in descending order

\begin{equation*}
    \lambda_1>\lambda_2>\dots>\lambda_n,
\end{equation*}
we define the relative spectral gap ($SG_r$) as

\begin{equation}
\label{eqn:SGr}
    SG_r=\frac{\lambda_1-\lambda_2}{\lambda_1}.
\end{equation}
The aforementioned definition of the relative spectral gap can be extended to monopartite directed networks substituting \textbf{A} to \textbf{B}.

\subsection{Artificial bipartite networks with high spectral gap}
In their general structure, networks with high spectral gap (typically $SG_r\ge 0.40$~\cite{benzi2013ranking}) present: (i) high connectivity among their nodes~\cite{tanner1984explicit,estrada2006spectral,alon1986eigenvalues,petri2013topological,estrada2007topological,jun2010natural,hoory2006expander,milanese2010approximating} and (ii) high correlation values among several centrality measures~\cite{estrada2006network,oldham2019consistency,benzi2013ranking,benzi2015limiting,mihail2002eigenvalue}. 
Aiming to explore the correlation among different centrality metrics in high spectral gap networks, we artificially generated an ensemble of bipartite networks having this peculiar characteristic. We considered two classes of graphs: (i) networks with pseudo-triangular incidence matrix, that we define as Pseudo-Triangular Networks (PTNs)~\cite{mariani2019nestedness,battiston2014metrics}, and (ii) Bipartite Erd\H{o}s-R\'enyi Graphs (BERGs)~\cite{erdHos1959random,newman2018networks}. 

A bipartite network is defined as a PTN (Pseudo-Triangular Network) if by permuting the rows and columns of its associated incidence matrix \textbf{B}, it is possible to identify a separatrix (i.e., a border-line) above which the density of links -- and, hence, of the non-zero values -- is much higher than below it~\cite{mariani2019nestedness}.  
The separatrix is ideally placed in correspondence of the diagonal of the incidence matrix~\cite{mariani2019nestedness}. Instead, \emph{full triangular matrices} -- and corresponding networks -- are those matrices in which, with a proper permutation of rows and columns, all the non-zero values lie above the separatrix, while all the zero ones are below it.
The matrix describing the international trade of goods~\cite{tacchella2012new,wu2016mathematics,battiston2014metrics} in a given year is an example of PTN. Other examples include the plant-pollinator system in ecology~\cite{james2012disentangling,bascompte2003nested, staniczenko2013ghost}, the interbank payment network~\cite{soramaki2007topology} and the manufacturer-contractor network~\cite{saavedra2009simple}. 

\begin{figure*}
\centering
\includegraphics[width=\textwidth]{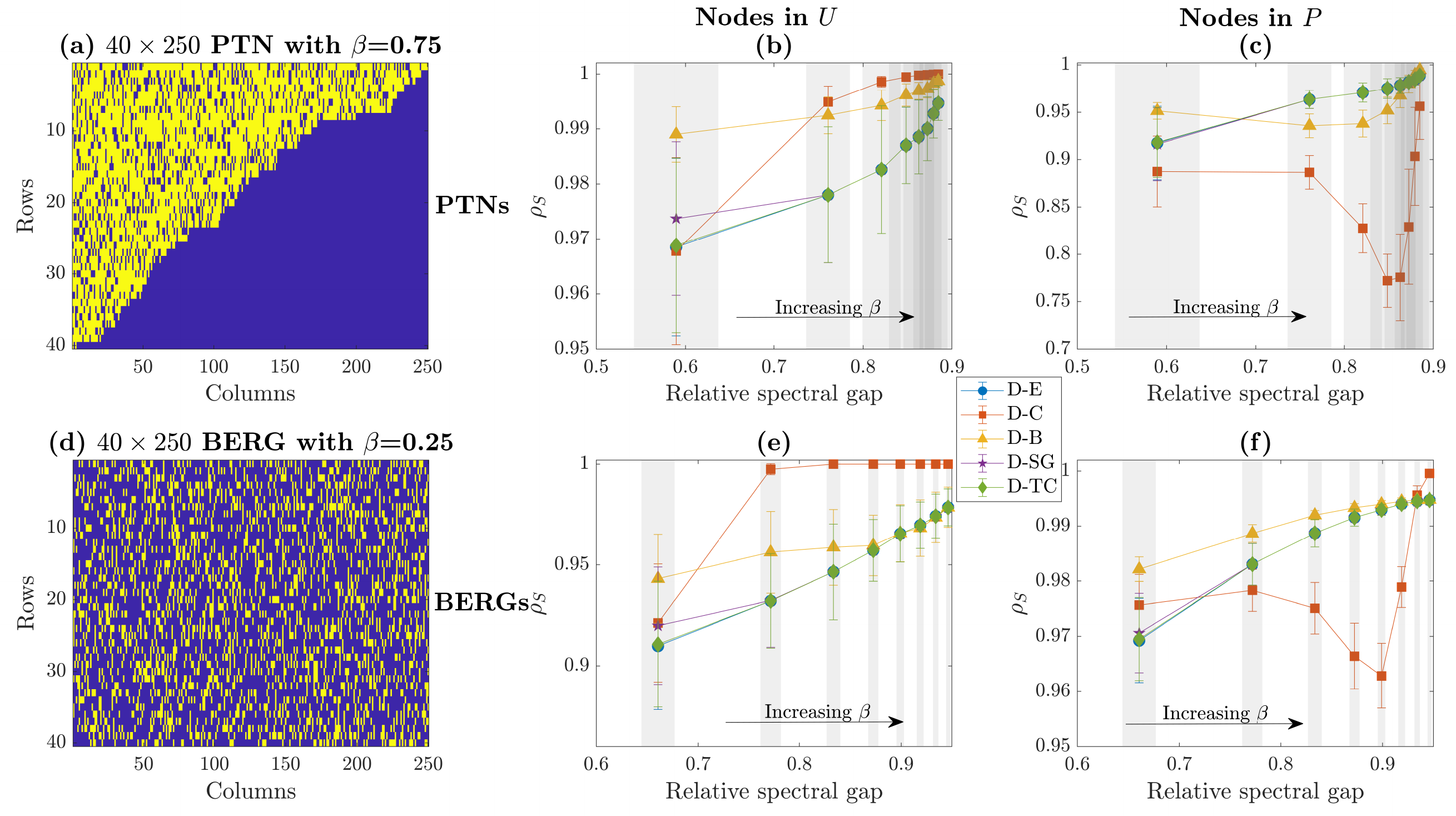}
\caption{Examples of two incidence matrices for each class of artificially generated networks of dimension $40 \times 250$ with a specific value of the parameter $\beta$ (yellow entries are nonzero values and blue otherwise) and corresponding correlation values between the degree and other centrality metrics as applied to the set of equally sized networks. The Spearman’s correlation coefficients ($\rho_S$) between the degree (D) and the eigenvector (E), closeness (C), betweenness (B), subgraph (SG) centrality, and total communicability (TC) are defined as functions of the relative spectral gap, and the computation of the metrics for both nodes in the $U$ and $P$ sets of the networks is detailed. Top panels refer to Pseudo-Triangular Networks (PTNs), while bottom ones to Bipartite Erd\H{o}s-R\'enyi Graphs (BERGs). In panels (b), (c), (e) and (f), each point represents the mean among $N_{sim} = 100$ networks of the considered class, the whiskers and the shaded regions describe $\pm1$ standard deviation of the correlation values and relative spectral gap, respectively. The arrows indicate the direction in which the threshold values $\beta$ increase for both PTNs and BERGs.}
\label{fig:problem}
\end{figure*}

For each class, we generated an ensemble of $N_{sim}=100$ networks according to the following two procedures. Notice that all generated networks are unweighted (i.e., the weights of the links are all unitary); moreover, in the generation of the networks, we only considered connected graphs. 

In defining each network within the PTNs ensemble, once the system's dimension has been set (i.e., the number of nodes), we started by generating a full triangular incidence matrix, which is then emptied. For each link (i.e., $B_{ij}=1$), a random edge probability was extracted from a uniform distribution between $0$ and $1$. The link at hand was maintained if the extracted probability was smaller than a given threshold value $\beta$, and removed otherwise. In this way, Pseudo-Triangular Networks defined in a wide range of densities were obtained as a function of the threshold value $\beta$. 

Instead, for the Bipartite Erd\H{o}s-R\'enyi Graphs (BERGs) ensemble, starting from an empty matrix with fixed dimensions, we filled the matrix with entries. Using a uniform distribution between $0$ and $1$, we extracted a random value defining the probability of a link between the node $i$ of set $U$ and $j$ of set $P$. A network was generated considering the links with a probability value smaller than the threshold value $\beta$. In this way, also the Bipartite Erd\H{o}s-R\'enyi Graphs are defined in a wide range of densities, obtained as a function of the threshold value $\beta$. 

In setting the dimensions of the networks, the elements in $U$ range in the interval $[10, 100]$, with step $15$, while those in $P$ in the interval $[100, 500]$ with step $25$, thus totalling $119$ dimension combinations for the generated artificial network systems. 
In generating the networks, the link threshold value $\beta$ for PTNs ranges in the interval $[0.15, 0.85]$ with step $0.10$; while for BERGs networks, it varies in the range $[0.10, 0.45]$ with step $0.05$. The $\beta$ values are defined in different ranges to generate artificial networks with comparable relative spectral gaps. 

These two ensembles define the experimental context wherein analyzing the relationship between spectral gap and the occurrence of correlation among centrality metrics. For each system's dimensions, the PTNs and BERGs ensembles can be analyzed through the perspective of the threshold value $\beta$, determining the density of the networks. In the main text, all results are shown for the ensemble of artificial networks of size $40 \times 250$, but similar evaluation and comments arise from the analysis of other network sizes (see Supplemental Material~\cite{supplementary}, figures S1 and S2). 

Figure~\ref{fig:problem} exemplifies two of the generated artificial networks through their incidence matrices, describing a Pseudo-Triangular Network in panel (a), and a Bipartite Erd\H{o}s-R\'enyi Graph one in panel (d). 
Figure~\ref{fig:problem} also shows the Spearman's correlation coefficients between the degree and other centrality metrics (eigenvector, closeness, betweenness, subgraph and total communicatility) obtained from the analysis of all generated matrices of size $40 \times 250$ (the Spearman's correlation metric is a ranking-based one). The correlation is given as a function of the relative spectral gap. As the figure shows, the two ensembles of networks are defined within comparable -- although not equal -- intervals of relative spectral gap. Moreover, the relative spectral gap is high~\cite{friedman1991second,staniczenko2013ghost,feng2014heterogeneity} and it increases as the threshold value $\beta$ increases, thus with increasing number of connections in the networks~\cite{jun2010natural} for both PTNs and BERGs. In general, for both network classes, the Spearman's correlation coefficient takes high values, with increasing trend as the relative spectral gap increases (in line with previous studies~\cite{oldham2019consistency,benzi2015limiting}).
Therefore, although being global centrality measures, in terms of nodes' ranking the eigenvector, closeness, betweenness, subgraph and total communicability centrality measures add little information compared to that provided by the degree centrality, in both network cases. This result can be further interpreted considering the existing literature. Benzi et al.~\cite{benzi2015limiting} demonstrated that the rankings provided by several centrality measures converge to the ones obtained by applying the degree or eigenvector centrality for high spectral gap networks. As shown in Figure~\ref{fig:problem} (as well as in the Supplemental Material~\cite{supplementary}, figures S1 and S2), the rankings obtained from these two metrics also present extremely high Spearman's correlation coefficients. Thus, following the guidelines by Li et al.~\cite{li2015correlation}, it is possible to approximate the more time consuming centrality, i.e., the eigenvector one, to simpler one, i.e., the degree, if the correlation between the two metrics is high. Therefore, for the networks at hand, the degree centrality dominates the centrality landscape. 

\section{\label{sec:GENEPY}Beyond the degree: the GENEPY index}
\subsection{Definition}
The Generalized Economic Complexity index (GENEPY) was introduced by Sciarra et al.~\cite{sciarra2020reconciling} to reconcile the contrasting methodologies within the field of Economic Complexity. It is grounded on the statistical and multidimensional approach to network centrality which is described in~\cite{sciarra2018change}. The methodologies of Economic Complexity were originally proposed to update the simpler measure of export complexity of countries provided by the degree~\cite{hidalgo2009building,tacchella2012new}. Following up on the idea of overcoming the limitation of the degree in the economic context, we introduce here the use of the GENEPY to the more general context of high spectral gap networks. 
The description of the GENEPY framework applied to a general network structure follows, and we refer the reader to~\cite{sciarra2020reconciling} for further technical details.

Starting from the incidence matrix of a bipartite network \textbf{B}, the GENEPY index is defined by the introduction of a transformation matrix \textbf{W}, defined as
\begin{equation}
    W_{ij}=\frac{B_{ij}}{k_ik'_j},
    \label{eqn:W}
\end{equation}
where 
$k_i=\sum_{j=1}^{N_P} B_{ij}$ and 
$k'_j=\sum_{i=1}^{N_U} \frac{B_{ij}}{k_i}$.
\textbf{W} preserves the network topology \textbf{B} describes, with the advantage of partially filtering away the degree-biased information thanks to the division of the $B_{ij}$ entries for $k_ik_j'$~\cite{costantini2021}.  
The GENEPY index of a generic node in either two sets $U$ and $P$ is the result of the combination of the two largest eigenvalues ($\phi_1$ and $\phi_2$, with $\phi_1>\phi_2$) and corresponding eigenvectors ($\textbf{x}_1$ and $\textbf{x}_2$) of a \emph{proximity matrix} defined from \textbf{W}. For the nodes in $U$, the proximity matrix \textbf{N} is defined as
\begin{equation}
    N_{ii^*}=
    \begin{cases}
    \sum_j W_{ij}W_{i^*j} &\text{ for } i\neq i^*, \\
    0 &\text{ for } i=i^*;
    \end{cases} \\ 
    \label{eqn:N}
\end{equation}
whereas, for the ones in $P$, the proximity matrix \textbf{G} is defined as
\begin{equation}
    G_{jj^*}=
    \begin{cases}
    \sum_i W_{ij}W_{ij^*} & \text{ for } j\neq j^*, \\
    0  & \text{ for } j=j^*.
    \end{cases}
    \label{eqn:G}
\end{equation}

The GENEPY index of a node $i$ is defined as 
\begin{equation}
\label{eqn:GENEPY}
    GENEPY(i)=\Bigl(\sum_{t=1}^2 \phi_{t}\text{x}_{t,i}^2\Bigr)^2+2\sum_{t=1}^2 \phi_t^2\text{x}_{t,i}^2,
\end{equation}
where $\text{x}_{t,i}$ is the $i$-th component of the eigenvector at the dimension $t=1,2$ of the proximity matrix at hand (either \textbf{N} or \textbf{G}). Since \textbf{N} and \textbf{G} are symmetric square matrices, their eigenvectors are distinctly defined (i.e., the right and left eigenvectors are the same). 

The GENEPY index per each set of nodes is derived from the matrices \textbf{N} and \textbf{G} describing the similarities among the connectivity pattern of the nodes in the network. Such patterns are differently interpreted by each proximity matrix, and to compute the GENEPY index for the nodes in $U$ or for those in $P$ provides different information regarding the centrality of the nodes~\cite{costantini2021}. 
Specifically, the GENEPY computed onto the proximity matrix \textbf{N} ranks the nodes according to two criteria: (i) the number of common neighbors that the considered node shares with other nodes in the network, and (ii) how many neighbors of the node at hand are also in the neighborhood of only the top raked nodes (by the GENEPY index computed onto \textbf{N}). Instead, the GENEPY computed onto \textbf{G} ranks first those nodes that share some neighbors with many other nodes in the network and are also connected to the more isolated nodes. Therefore, we expect the GENEPY computed onto \textbf{N} to be better correlated with the degree centrality, since the matrix \textbf{N} defines quantitative similarities in the number of connections of the nodes. Conversely, the matrix \textbf{G} defines structural similarities in the way the nodes are connected, and thus a lower correlation between the degree of the nodes and the GENEPY index from \textbf{G} is expected. To further show the functioning of the GENEPY index, we provide an example for the bipartite networks in the Appendix, and one considering the monopartite networks in the Supplemental Material~\cite{supplementary}.

It is worth to notice that the matrix \textbf{N} (\textbf{G}) is the one-mode projection of matrix \textbf{W} on $U$ ($P$), with the diagonal values set to $0$. 
To set the diagonal values to zero entails removing the information about the degree of the nodes, contained in the nominator through the values $\sum_j B_{ij} B_{i^*j}$ for $i^*=i$ (see eqs.~\eqref{eqn:W} -~\eqref{eqn:G}). Therefore, setting to $0$ these values filters away the information of the degree from the one-mode projection. 
Another crucial feature of the GENEPY is its multidimensional framework: the first two eigenvectors of the proximity matrices are combined in this metric. They carry different information: $\textbf{x}_\textbf{1}$ is the eigenvector centrality of the proximity matrix \textbf{N} (\textbf{G}) that scores the elements in $U$ ($P$) according to the similarities in their neighborhood (the set of nodes in $P$ connected to a specific node in $U$ and \emph{vice-versa} for the elements in $P$). In contrast, the second eigenvector $\textbf{x}_\textbf{2}$ identifies clusters of nodes in $U$ ($P$) according to their connections~\cite{sciarra2018change,sciarra2020reconciling,mealy2019interpreting,gfeller2007spectral,estrada2007topological}.
A final comment on $k'_j$ is due to the reader. This term describes the degree of an element $j$ in $P$ taking into account the degree and the number of nodes in $U$ to which $j$ is connected. In the transformation from \textbf{B} to \textbf{N} (or \textbf{G}, eqs.~\eqref{eqn:W} -~\eqref{eqn:G}), the term $\frac{1}{k'_j}$ quantifies how poorly connected is the element $j$ with the nodes in $U$~\cite{sciarra2020reconciling}. As a consequence, the elements of \textbf{N} (\textbf{G}) associated to a small value of $k'_j$ receive a higher weight. 
Therefore, the GENEPY ranks the nodes in $P$ from the least to the most central, thus returning an anti-centrality score. 
It follows that, to compare the nodes' importance given by the GENEPY index (anti-centrality measure) and the degree (centrality measure) for the nodes in $P$, the GENEPY rankings should be inverted (e.g., by multiplying each GENEPY value by $-1$).

\subsection{Tests on artificial networks}

\begin{figure*}
\centering
\includegraphics[width=\textwidth]{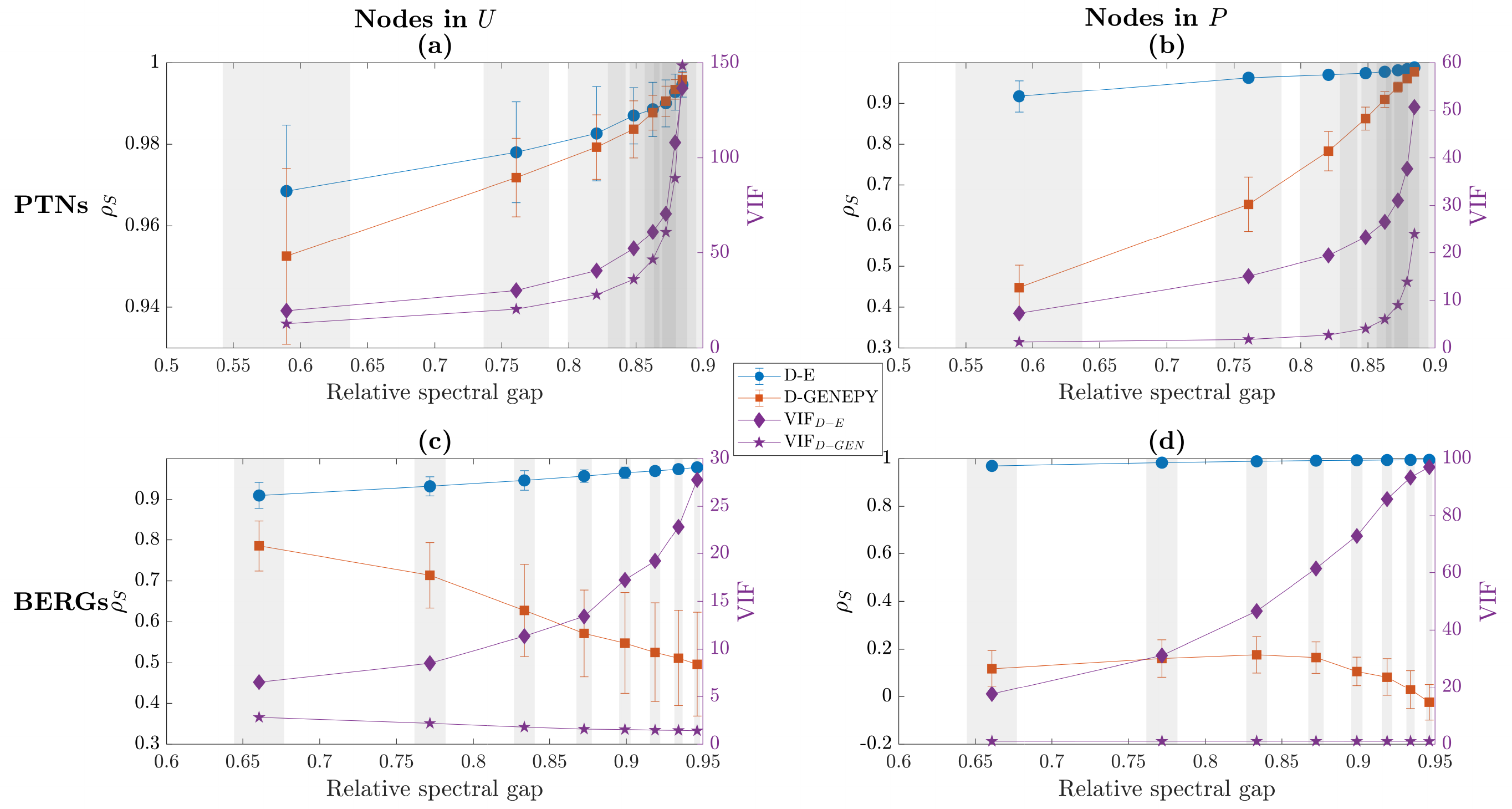}
\caption{The Spearman's correlation coefficients among different centrality measures for the Pseudo-Triangular Networks (PTNs), panels (a) and (b), and BERGs ones (Bipartite Erd\H{o}s-R\'enyi Graphs), panels (c) and (d). The correlation values ($\rho_S$) between the degree (D) and eigenvector (E) centrality (blue dots), and the degree and GENEPY index (red squares), are shown as functions of the relative spectral gap. The right panels refer to the $U$ set, the left ones to the $P$ set. Each point represents the average value of $N_{sim}=100$ realizations, and $\pm1$ standard deviations are given for both correlation (whiskers) and relative spectral gap (shaded regions) values. In purple, we show the mean Variance Inflation Factor (VIF) computed over the $N_{sim}$ artificial networks between the degree and eigenvector centrality (diamonds) and the  degree-GENEPY values (stars).}
\label{fig:GEN_res}
\end{figure*}

Following the correlation analysis onto the artificially generated bipartite networks, we also test the performance of the GENEPY index by computing the Spearman's correlation between the proposed metric and the degree, and by comparing it with the degree-eigenvector correlation for both Pseudo-Triangular Networks and Bipartite Erd\H{o}s-R\'enyi Graphs. Other correlation results between the GENEPY and the centrality metrics used in this work can be found in the Supplemental Material, figures S3-S7~\cite{supplementary}. 
Aiming to demonstrate the added value of the use of the GENEPY, we introduce here the Variance Inflation Factor (VIF)~\cite{mansfield1982detecting,daoud2017multicollinearity,alin2010multicollinearity}. The VIF measures the collinearity between two variables (or more than two, in which case the multicollinearity is measured~\cite{alin2010multicollinearity}). Collinearity quantifies whether the variable at hand is related through a linear relationship to another one, collinearity being significant for VIF values larger than $5$~\cite{daoud2017multicollinearity}. The VIF is defined as
\begin{equation}
    VIF=\frac{1}{1-\rho_S^2},
    \label{eqn_VIF}
\end{equation}
where $\rho_S$ is the Spearman's correlation coefficient between the variables at hand. 

We use the VIF to check the collinearity among the rankings obtained by applying the degree, eigenvector centrality and GENEPY, thus allowing one to quantify the distinct information the GENEPY captures with respect to the degree or eigenvector centrality. 

In figure~\ref{fig:GEN_res}, we plot the correlation coefficients among the considered centrality measures and the VIF values as a function of the relative spectral gap for PTNs and BERGs (top and bottom panels, respectively), and for both sets $U$ and $P$ (left and right panels, respectively). As the VIF values indicate, the use of the GENEPY in all considered networks (and nodes' sets) allows one to overcome the degree-dominated information. As a general rule of thumb, the VIF from the degree-GENEPY correlation, VIF$_{D-GEN}$, is always lower than the one computed on the degree-eigenvector correlation, VIF$_{D-E}$. These latter values increase with increasing spectral gap. Although showing the same increasing trends in case of Pseudo-Triangular Networks (panels 2a and 2b), the VIF$_{D-GEN}$ values stand below the VIF$_{D-E}$ ones. Stable trends as function of the relative spectral gap can be seen for the Bipartite Er\H{o}s-R\'enyi Graphs (panels 2c and 2d). In particular, in these cases the VIF$_{D-GEN}$ values are smaller than $5$, confirming that the GENEPY rankings are not collinear with those obtained by applying the degree; thus, due to their high correlation with the degree, neither it does with the other centrality metrics (see figure S3-S7 of the Supplemental Material~\cite{supplementary}). 

Looking at the distinction between the two sets of nodes, differences in the correlation values across the sets also emerge (see the left panels 2a and 2c for the $U$ nodes, the right ones, 2b and 2d, for the $P$ nodes), for both PTNs and BERGs cases. As expected, higher correlations between the degree and GENEPY are found for the nodes in $U$ rather than for those in $P$ (similar results were obtained for the other network sizes, whose outcomes are given in figures S8-S11 in the Supplemental Material~\cite{supplementary}). 

Notice that the choice of interpreting the nodes in $U$ and $P$ through the matrices \textbf{N} and \textbf{G}, respectively, is arbitrary in artificial networks since there is no physical meaning of the nodes in the simulations. In fact, we can reverse our interpretation of the system by reversing the analysis of the two sets of nodes, thus computing the GENEPY by transposing the incidence matrix \textbf{B} describing the artificial system at hand. Therefore, the sets of nodes $U$ and $P$ would be analyzed by means of the proximity matrices \textbf{G} and \textbf{N}, respectively. This reversed interpretation of the system cannot be considered in the cases of real-world bipartite systems, where the choice of mapping the $U$ and $P$ node sets in \textbf{N} or \textbf{G} depends on the nodes' significance in the system. In any case, as we show in figure S12 of the Supplemental Material~\cite{supplementary}, to reverse such analysis in the artificially generated networks does not change the results presented in figure~\ref{fig:GEN_res}. In particular, the GENEPY rankings computed on \textbf{N} and \textbf{G} are always collinear for Pseudo-Triangular Networks, while differs for Bipartite Erd\H{o}s-R\'enyi Graph. However, as shown in figure~\ref{fig:GEN_res}, the information provided by the GENEPY differs from the degree or eigenvector centrality. 

\subsection{Application to real-world networks}
In support of our results on artificially generated networks, we extend the spectral gap-VIF analysis to real-world networks, in both monopartite and bipartite systems. These tests allow one to further comprehend how the GENEPY captures complementary characteristics of the system under study with respect to previously proposed metrics. 

\begin{figure*}
\centering
\includegraphics[width=\textwidth]{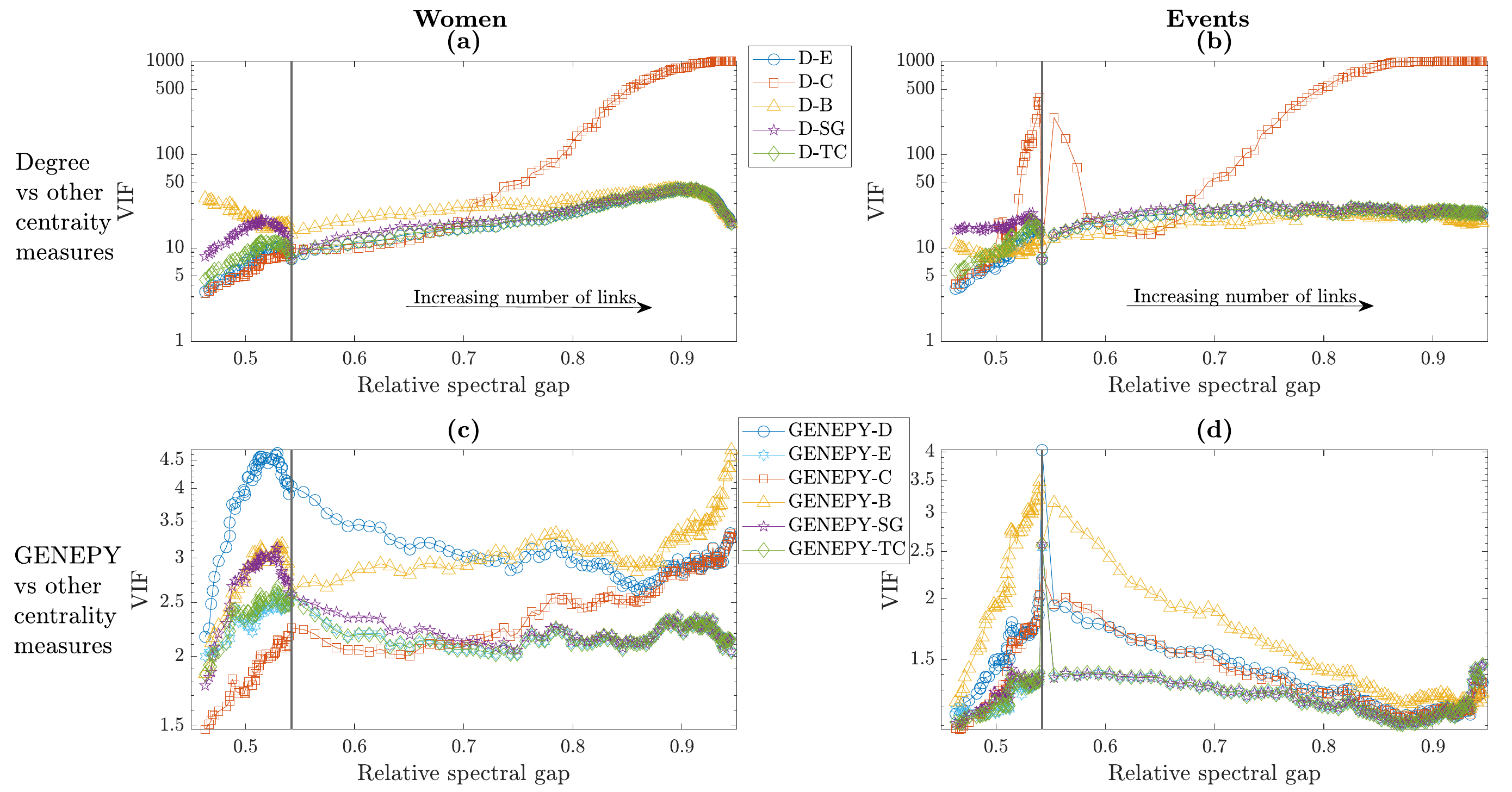}
\caption{Values of the Variance Inflation Factor computed among the centrality metrics (degree -- D --, eigenvector -- E, blue circles --, closeness -- C, red squares --, betweenness -- B, yellow triangles --, subgraph centrality -- SG, purple pentagrams -- and total communicability -- TC, green diamonds --) applied onto the Women-Events network~\cite{TwoModeData,davis2009deep} and its simulation with edges' addition/removal. The benchmark for the computation of the VIF values is the degree centrality in the top-panels (panel (a) for the women set, and panel (b) for the events set); the GENEPY index computed onto \textbf{N} in panel (c); the GENEPY index computed onto \textbf{G} in panel (d). To modify the spectral gap, $100$ ($50$) links were randomly added (removed), one at a time, to the original network. Each point is the mean, computed at the same number of links in the network, among $100$ repetition of the edges' addition/removal process. For the sake of graphical representation, the infinite values of the VIF (corresponding to a Spearman's correlation of $1$) were replaced by VIF$=1000$ (Spearman's correlation value $0.9995$). The black arrows indicate the direction in which the links were added, and the vertical line in each panel highlights the VIF values and relative spectral gap of the original network.}
\label{fig:women}
\end{figure*}

We start testing the GENEPY performances onto the well-known bipartite system "Women-Events network", detailing the attendance of women to social events in the late 30's \cite{davis2009deep,TwoModeData}. The system constitutes of $18$ women and $14$ events. The relative spectral gap of this network is quite high, $0.54$, and the rankings for both kind of entities computed according to the degree, eigenvector, closeness, betweenness, subgraph and total communicability centrality are collinear, i.e., their VIF values are greater than $5$ (see points of the plots in panels (a) and (b) of figure \ref{fig:women} indicated by the vertical black lines). Since the spectral gap is related to the level of connectivity of the network (as shown by Jun et al.\cite{jun2010natural} and confirmed by our simulations on artificially generated networks), we performed some numerical experiments to further investigate the relationship between centrality metrics and spectral gap. 
To this aim, we randomly added (and removed) $100$ ($50$) edges --one at a time-- in the incidence matrix of the network. We repeated the link addition/removal process $100$ times, thus generating an ensemble of simulations at different number of links in the so-modified network. Therefore, for each ensemble of equal number of links, we evaluated the mean values of the relative spectral gap and the VIF values among the centrality measures considered in this work. Figure \ref{fig:women} presents the results from these simulations for each set of nodes, and it details the VIF values among the degree, GENEPY and other centrality metrics. As the number of links increases, the relative spectral gap and the VIF values also increase, which is in line with the simulations on the artificially generated networks. In particular, the VIF values computed among the degree and the previously proposed metrics is in general greater than $5$. 

The comparisons between the GENEPY and the other metrics are presented in panel (c) for the women and (d) for the events. In both cases, the GENEPY is never collinear to other metrics, highlighting that the information it provides differs from that of the other centrality measures. 
In computing the GENEPY, the set $U$ details the women (thus the GENEPY is computed onto the proximity matrix \textbf{N}), while the set $P$ the events (i.e., the GENEPY computed onto the proximity matrix \textbf{G}). We can interpret the centrality ranking provided by the GENEPY considering its rationale and the system at hand: a woman is considered more central if she co-attends events participated with many other women, some of which exhibit high centrality. Instead, an event is classified as central when it is attended, at the same time, by both more and less socially engaged women, since the participation of high-centrality women increases its centrality score. The corresponding results in correlation are reported in figure S13 of the Supplemental Material~\cite{supplementary}.

Similar results hold for the monopartite example of the Zachary's karate club network~\cite{zachary1977information}. The results, and related comments, are reported in the Supplemental Material, figures S14 and S15~\cite{supplementary}.

\begin{figure*}
\centering
\includegraphics[width=\textwidth]{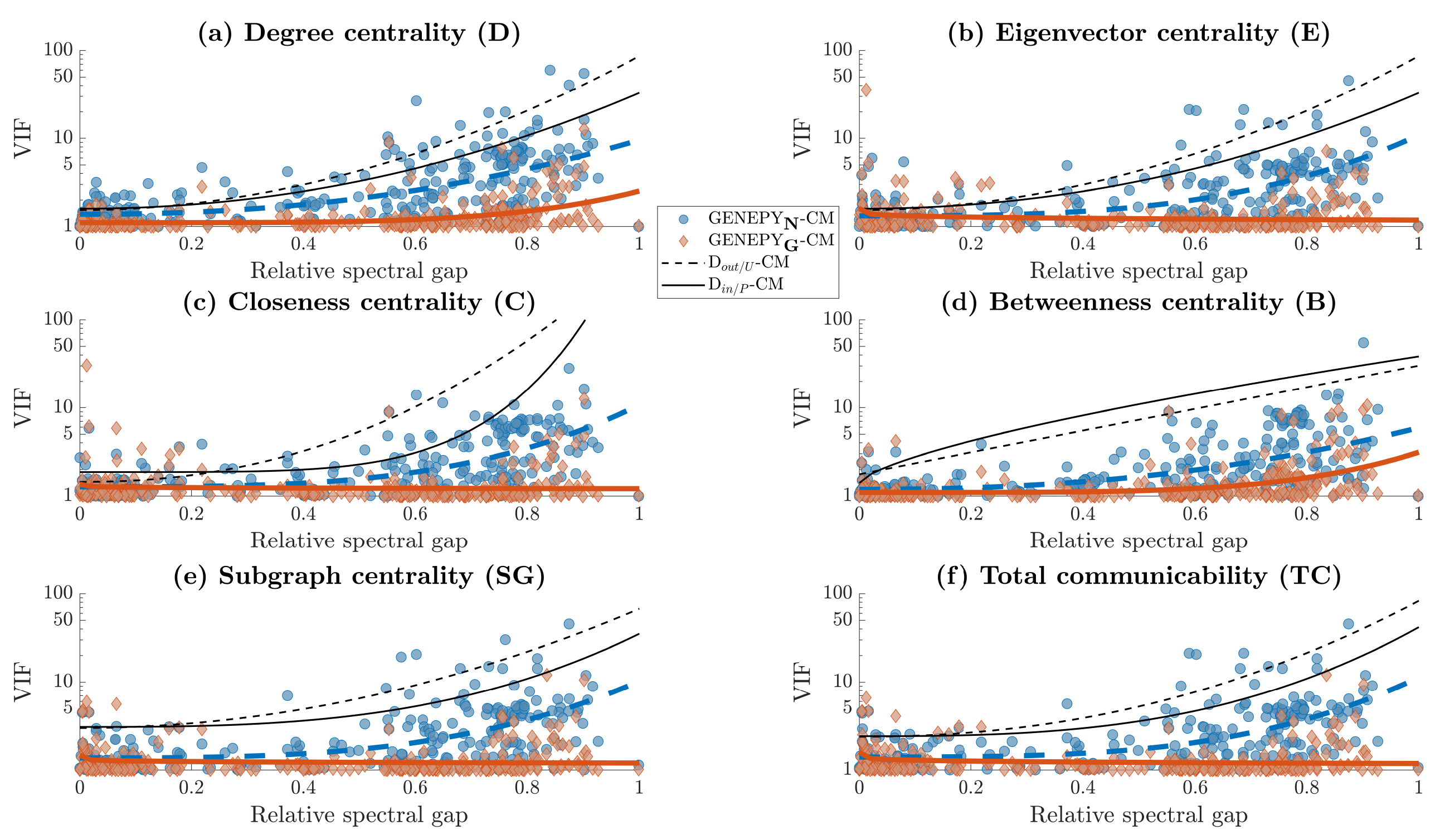}
\caption{Comparison of the VIF values between the GENEPY and the degree (D, panel (a)), eigenvector (E, panel (b)), closeness (C, panel (c)), betweenness (B, panel (d)), subgraph (SG, panel (e)) and total communicability (TC, panel (f)) centrality measures as a function of the relative spectral gap for the $284$ real-world networks considered in this work. In each panel, the blue circles (red diamond) report the VIF values computed between the GENEPY calculated onto the proximity matrix \textbf{N} (\textbf{G}) with the Centrality Measure (CM) reported in the panel title. The thick blue dashed (thick red solid) lines fit the logarithm (base $10$) of blue circles (red diamonds) with a function $a(SG_r)^b+c$. 
The thin dashed and solid black lines describe the fits of VIF values between the degree centrality and the centrality measure reported in the title of each panel. For sake of graphical representation, the y-axis is limited to the range $[0, 100]$ to enhance the readability of the points.} 
\label{fig:Real_Cases}
\end{figure*}

The results on the spectral gap, VIF and correlation obtained from the Women-Events and Zachary networks are fully confirmed by the tests run on a wide set of real-world network. We considered three different sets of systems: (i) the international trade of goods, taken from the BACI-CEPII dataset~\cite{gaulier2010baci}; (ii) the ecological systems, from the Web of Life dataset (www.web-of-life.es); and (iii) monopartite systems (both undirected and directed) from the SuiteSparse Matrix Collection~\cite{davis2011university} (https://sparse.tamu.edu). The variety of these real-world systems allows us to extend the analysis of the GENEPY performances on a broader class of networks, also extending the proposed centrality onto monopartite graphs. 

For the sake of comparison with the artificially generated networks, we only considered the networks' topology, thus neglecting any links' weights. In our analysis, we included all connected networks with at least seven nodes in the system. As a result, a total of $284$ 
networks were considered: $23$ from the trade systems (years from $1995$ to $2017$, the nodes in $U$ are in the range $[174, 182]$ and those in $P$ are between $1201$ and $1241$), $126$ ecological networks (nodes in $U$ in the interval $[7, 456]$ while those in $P$ are in the range $[7, 1044]$) and $135$ monopartite networks (of which, $104$ undirected and $31$ directed, with number of nodes between $8$ and $6765$). The results of these tests are shown in figure~\ref{fig:Real_Cases} (the corresponding outcomes in correlation values are in figure S16 of the Supplemental Material~\cite{supplementary}).

For each centrality metrics, the figure shows their VIF values with the GENEPY. The VIF behavior as functions of the relative spectral gap is coherent with what observed for the synthetic graphs and the Women-Events network: the GENEPY index ranks the nodes differently from the metrics previously proposed in network theory. 
This is also confirmed by the fitting lines reported in the figure; in fact the thick coloured lines fitting the VIF values between the GENEPY and the considered centrality metric are always below the thin black lines fitting the degree and the metric indicated in the panel title. These results endorse the outcomes from the analysis on artificial networks about the potential of the GENEPY framework in changing the perspective with respect to the other centrality measures.

Since the plots in figure~\ref{fig:Real_Cases} refer to both undirected and directed networks, a further comment is needed to understand the application of the GENEPY framework to monopartite graphs. 
The GENEPY index provides different results whether the matrix \textbf{N} or \textbf{G} is used for computation. In particular, due to the similarities of bipartite graphs with directed ones~\cite{newman2018networks}, in interpreting the result of the GENEPY framework we associate the analysis related to the set $U$ (i.e. the GENEPY onto \textbf{N}) with the out-going properties of the nodes, while the one for the set $P$ (i.e. the GENEPY onto \textbf{G}) with the in-coming properties. 

As shown in figure~\ref{fig:Real_Cases}, the lines fitting the VIF values between the degree and the usual centrality measures increase as the the relative spectral gap increases. In particular, over a certain value of the spectral gap, any centrality measure present collinearity (VIF larger than $5$) with the degree. This fact is in line with the conclusion of Benzi et al.~\cite{benzi2015limiting} for the degree, eigenvector, subgraph and total communicability centrality measures. In this work, we have enlarged these observations to the closeness and betweenness centrality (see thin black lines in panels 4c and 4d, respectively). Instead, the VIF values between the GENEPY and the other metrics exhibit a different behavior whether they are computed onto \textbf{N} or \textbf{G}, in any case maintaining a non collinear nodes' ranking with the other metrics. 

The VIF values of the GENEPY computed on \textbf{N} are typically larger than those computed on \textbf{G}. This is consistent with our previous comments on the rationale behind the GENEPY computed on the proximity matrices \textbf{N} and \textbf{G}. In fact, because the GENEPY on \textbf{N} is related to the number of neighbors a node shares with the others, it is more correlated with the degree (and thus with all the other metrics) than the GENEPY computed on \textbf{G}. Furthermore, the fits of the VIF values corresponding to GENEPY on \textbf{N} (thick dashed blue lines) exhibit an increasing trend as the spectral gap increases, whereas the behavior of the fits of the VIF values due to the GENEPY on \textbf{G} (thick solid red lines) depends on the centrality metric against which the GENEPY is compared. The solid thick lines increase when the GENEPY on \textbf{G} is compared with the degree and betweenness centrality, while it remains nearly constant for the other centrality measures. This confirms the fact that GENEPY provides two different centrality perspectives on the nodes in the network in question that are less affected by the spectral gap than the other centrality measures.

\section{\label{sec:conclusion}Conclusions}
In this work, we propose to use the GENEPY index to shed new light on high spectral gap networks. The proposed metric ranks the nodes according to the similarities in their connectivity pattern. We compared the GENEPY performances with other commonly used centrality measures (as the degree, eigenvector, closeness and betweenness centrality) onto both synthetic and real-world networks. 
As we have shown, the previously proposed metrics present high correlation and collinearity with the degree centrality for increasing relative spectral gap. Conversely, the GENEPY metrics and the other centrality measures are less collinear, entailing that the GENEPY can unveil new centrality characteristics of the nodes. 

The outcomes of our work demonstrate that the GENEPY centrality offers a complementary approach to the network centrality problem than those previously proposed, which means that the GENEPY has the capability to enlarge the information that can be obtained from an interacting system. 

The rankings' correlation analysis is one of the most direct ways to evaluate the differences among several centrality measures, yet other approaches can be employed. For example, trending approaches consider centrality metrics as the explanatory variable in a prediction exercise, where the aim is to proxy a target variable characterizing the system at hand through suitably system-specific functions (e.g.,~\cite{hartmann2017linking,albert2007network}). Our results pave the way for future research aimed to compare the GENEPY and other centrality metrics (in high spectral gap systems) from the predictive capability perspective.

\appendix

\section{A toy model to explain how the GENEPY works}
\begin{figure}
    \centering
    \includegraphics[scale=0.45]{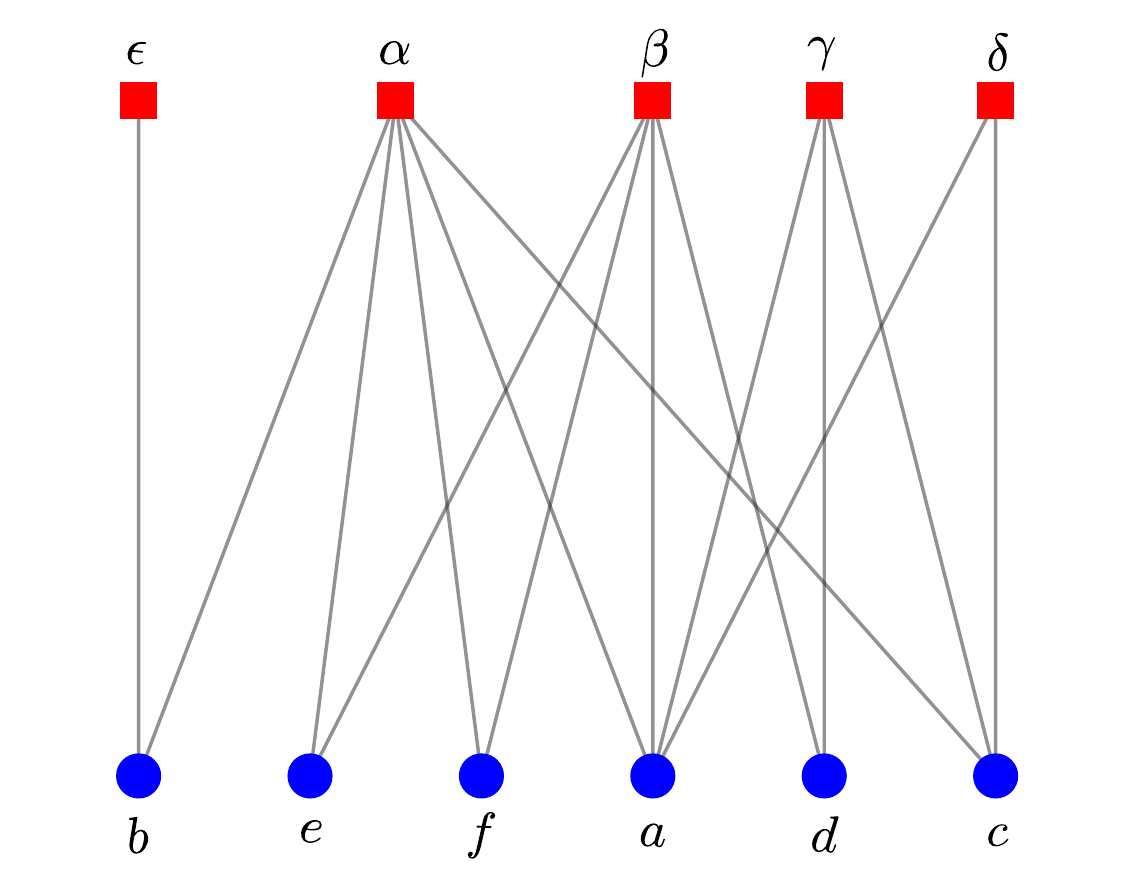}
    \caption{Representation of the bipartite network used as toy model. The nodes in the set $U$ (indicated with the Greek letters $\alpha$, $\beta$, $\gamma$, $\delta$ and $\epsilon$) are represented by red squares, while those in $P$ (lowercase Latin letters $a$, $b$, $c$, $d$, $e$ and $f$) by blue circles. The black solid lines represent the connections among the nodes in the two sets. The nodes in $U$ are ranked according to the GENEPY computed onto \textbf{N}, while those in $P$ onto the proximity matrix \textbf{G}.} 
    \label{fig:appendix}
\end{figure}

To show how the GENEPY works, let us consider the bipartite network presented in Figure~\ref{fig:appendix} and described by the following incidence matrix: 

\begin{equation*}
    \textbf{B}=
    \begin{blockarray}{ccccccc}
    & a & b & c & d & e & f \\
    \begin{block}{c@{\hspace*{10pt}}(cccccc)}
    \alpha & 1 & 1 & 1 & 0 & 1 & 1 \\
    \beta & 1 & 0 & 0 & 1 & 1 & 1\\
    \gamma & 1 & 0 & 1 & 1 & 0 & 0 \\
    \delta & 1 & 0 & 1 & 0 & 0 & 0 \\
    \epsilon & 0 & 1 & 0 & 0 & 0 & 0 \\
    \end{block}
  \end{blockarray}
\end{equation*}


This network has a relative spectral gap of $0.77$. Aiming at showing the benefits of the proposed metrics, we compare the rankings obtained by applying to this graph the GENEPY index and the degree, eigenvector, closeness, betweenness, subgraph and total communicability centrality measures. 

In Tables~\ref{tab:Uranking} and~\ref{tab:Pranking}, one can see that the rankings obtained from the classical centrality measures (degree, eigenvector, closeness, and betweenness) and those from the eigen-properties of the incidence matrix (subgraph centrality and total communicability) are similar to each other. All metrics agree in ranking the nodes $\alpha$ and the node $a$ for the sets $U$ and $P$, respectively, as the most central ones in the network. This fact remarks that different approaches to solve the centrality problem onto the same network, provide about the same information if the spectral gap of the graph is high, as in this case. In contrast, the GENEPY index provides a different nodes' ranking based on the similarities among the connectivity pattern of the nodes, which is described by the proximity matrices \textbf{N} and \textbf{G} in a different way. Therefore, to compute the GENEPY index onto \textbf{N} and \textbf{G} provides different information regarding the centrality of the nodes. Specifically, the GENEPY computed onto the proximity matrix \textbf{N} considers two aspects of the nodes' connectivity pattern: (i) important nodes have one or more common neighbor/s (in $P$) with many nodes of the same set (i.e., $U$), and (ii) relevant nodes share common neighbors only with other important nodes. 
Considering these two aspects, the node $\beta$ is the most important element in $U$, according to the GENEPY centrality, because it is connected to the node $a$ in $P$, as the nodes $\alpha$, $\gamma$ and $\delta$, and shares neighbors with nodes $\alpha$ and $\gamma$ in an exclusive way (nodes $e$, $f$ and $d$, respectively). The node $\alpha$ is in the second position of the GENEPY ranking despite having the same number or exclusive connections of the node $\beta$. This is due to the fact that the node $\alpha$ shares the node $b$ as common neighbor with the node $\epsilon$, that is the least central element in $U$ for the GENEPY. Thus, for the GENEPY onto \textbf{N}, the node $\alpha$ loses relevance in the network and it is below the node $\beta$ in the ranking. 
Conversely, the GENEPY centrality computed onto the proximity matrix \textbf{G} ranks as relevant those nodes that, at the same time, share some neighbors with other nodes (in the same set) and are connected to some more isolated nodes of the other set. Thus, the node $b$ in $P$ is ranked as the most important node; in fact, the node $b$ is connected to the nodes $\alpha$ (as other nodes in $P$) and $\epsilon$, which is the most isolated node (according to the GENEPY metric).

\begin{table}
    \centering
    \caption{Comparison among the rankings (R) of the nodes in $U$ obtained by applying several centrality measures: degree (D), eigenvector (E), closeness (C), betweenness (B), subgraph centrality (SG), total communicability (TC) and the GENEPY computed onto \textbf{N} (GEN$_{\textbf{N}}$). The rankings are arranged starting from the most central node (with R=1) to the least central node (with R=5). The value in the brackets indicate the score obtained by applying the considered centrality measure.}
    \begin{tabular}{cccccccc}
    \toprule
        R & D & E & C & B & SG & TC & GEN$_{\textbf{N}}$ \\
    \hline
        1 & $\alpha(5)$ & $\alpha$(0.64) & $\alpha$(0.06) & $\alpha$(22.7) & $\alpha$(5.98) & $\alpha$(33.0) & $\beta$(0.73)  \\
        2 & $\beta$(4) & $\beta$(0.53) & $\beta$(0.05) & $\beta$(7.8) & $\beta$(4.71) & $\beta$(26.7) & $\alpha$(0.64) \\
        3 & $\gamma$(3) & $\gamma$(0.43) & $\gamma$(0.045) & $\gamma$(4.03) & $\gamma$(3.69) & $\gamma$(22.1) & $\gamma$(0.29) \\
        4 & $\delta$(2) & $\delta$(0.34) & $\delta$(0.041) & $\delta$(0.5) & $\delta$(2.73) & $\delta$(17.1) & $\delta$(0.16) \\
        5 & $\epsilon$(1) & $\epsilon$(0.07) & $\epsilon$(0.03) & $\epsilon$(0.0) & $\epsilon$(1.60) & $\epsilon$(5.43) & $\epsilon$(0.03) \\
    \hline
    \end{tabular}
    \label{tab:Uranking}
\end{table}

\begin{table}
    \centering
    \caption{Comparison among the rankings (R) of the nodes in $P$ obtained by applying several centrality measures: degree (D), eigenvector (E), closeness (C), betweenness (B), subgraph centrality (SG), total communicability (TC) and the GENEPY computed onto \textbf{G} (GEN$_{\textbf{G}}$). The rankings are arranged starting from the most central node (with R=1) to the least central node (with R=6). The value in the brackets indicate the score obtained by applying the considered centrality measure. Notice that, for the GENEPY index, the negative scores are given by the recast of the index into a centrality metrics, rather than an anti-centrality one.}
    \begin{tabular}{cccccccc}
    \toprule
       R & D & E & C & B & SG & TC & GEN$_{\textbf{G}}$ \\
    \hline
        1 & $a$(4) & $a$(0.62) & $a$(0.059) & $a$(10.7) & $a$(5.21) & $a$(30.8) & $b$(-0.019)  \\
        2 & $c$(3) & $c$(0.45) & $c$(0.053) & $b$(9) & $c$(3.75) & $c$(22.9) & $c$(-0.186) \\
        3 & $b$(2) & $e$(0.37) & $e$(0.048) & $c$(5.6) & $e$(2.86) & $e$(18.9) & $a$(-0.318) \\
        4 & $d$(2) & $f$(0.37) & $f$(0.048) & $e$(1.77) & $f$(2.86) & $f$(18.9) & $d$(-0.349) \\
        5 & $e$(2) & $d$(0.30) & $b$(0.044) & $f$(1.77) & $d$(2.62) & $d$(15.8) & $e$(-0.701) \\
        6 & $f$(2) & $b$(0.23) & $d$(0.040) & $d$(1.17) & $b$(2.43) & $b$(13.2) & $f$(-0.701) \\
    \hline
    \end{tabular}
    \label{tab:Pranking}
\end{table}


\providecommand{\noopsort}[1]{}\providecommand{\singleletter}[1]{#1}%

\end{document}